\documentclass[oldversion]{aa} 
\usepackage{graphicx}
\usepackage{txfonts}
\begin{document} 
\titlerunning{A deep catalog of morphologies of Coma galaxies}
\title{Morphology of galaxies in the Coma cluster region down 
to $\mathrm{M_B}=-14.25$}
\subtitle{I. A catalog of 473 members
\thanks{Based on observations obtained at the Canada-France-Hawaii Telescope 
(CFHT) which is operated by the National Research Council of Canada, the Institut 
National des Sciences de l'Univers of the Centre National de la Recherche Scientifique
of France, and the University of Hawaii} 
}
\author{R. Michard 
\inst{1}
\and S. Andreon
\inst{2}
}
\institute{
Observatoire de Paris, LERMA, 77 av. Denfert-Rochereau,
F-75015, Paris, France 
\and
INAF--Osservatorio Astronomico di Brera, via Brera 28, 20121, Milano, 
Italy\\
\email{stefano.andreon@brera.inaf.it}
}
\date{Received date; accepted date} 
\abstract
{This paper presents morphological type,
membership, and $U-V$ color for a sample of
galaxies in the Coma cluster direction, complete down to $M_B=-15.00$ mag
and extending down to $M_B=-14.25$ mag.
We have examined 1155 objects from the GMP 1983 catalog on B and V
images of the CFH12K camera, and obtained the Hubble type in most cases. 
Coma cluster membership for 473 galaxies was derived using morphology, 
apparent size, and surface brightness, and, afterward, redshift. 
The comparison among morphology- and redshift- memberships and among
luminosity functions derived from this morphologically-selected sample, or
by using statistical members or spectroscopic members, all show that the
morphological membership provided here can be trusted. 
For the first time, the morphological classification of Coma galaxies
reaches magnitudes that are faint enough to observe the whole magnitude 
range of the giant types, E, S0, and spiral stages. 
The data presented in this paper makes our sample the
richest environment where membership and morphology for complete samples 
down to faint magnitudes ($M_B\sim-15$ mag) are available,
thereby enlarging the baseline of environmental studies.}

\keywords{
Galaxies: luminosity function, mass function -- 
Galaxies: elliptical and lenticular, cD --
Galaxies: Spirals Galaxies: dwarf 
Galaxies: clusters: individual: Coma}

\maketitle

\section{Introduction} 
Morphological classification is probably the oldest tool used in galaxy studies, as
recently reviewed in Sandage (2005).  
Galaxies populate only a limited number of forms. With these forms are associated,  
or correlated, a number of such important properties as luminosities, internal dynamics,
and material composition, i.e. stars and ISM. In the study of such systems of galaxies 
as clusters, one would like to obtain statistically significant data, including of
course the morphological types of all objects. But as emphasized by Buta (2000) 
the art of morphology is strongly dependent on the resolution of the images of galaxies 
that span such a large domain of intrinsic sizes and distances. The 
{\em nec plus ultra} in galaxy morphology still is the catalog of 
Binggeli et al. (1985 or BST2) for the Virgo cluster. An important novelty of this
work was the use of morphology, apparent size, and surface brightness to ascertain 
the membership of faint galaxies in the cluster. It is indeed not feasible to extend 
redshift measurements to all detected objects, and the addition of appearance 
as a tool of distance estimates is welcome. The observation of new radial velocities 
(Binggeli et al. 1993) then confirmed nearly all the 1985 results.
 
The Coma cluster is one of the best-studied such objects because of its relative
proximity and because it may be a prototype of dynamically relaxed clusters, 
although significant substructures are present (Colless \& Dunn, 1996, Biviano et
al. 1996). This is in marked contrast to other clusters such as Virgo with its 
complex geometry and important substructure (Binggeli \& Huchra, 2000). 
Compared to the fair amount of photometric, 
colorimetric, and spectrographic information available for Coma cluster galaxies, 
morphological data are far from complete for these. 
Dressler (1980a, 1980b) described the overpopulation of early-type, E, and S0, galaxies 
in the Coma and other clusters in relation to their density, and introduced the 
morphology-density relation that is also discussed by Whitmore et al. (1993) and many
other later works. These later studies used 
detailed photometry to refine the classification of relatively bright objects 
(Jorgensen \& Franx, 1994; Michard, 1995; Andreon et al. 1996, 1997; Andreon 1996; 
Guttierez et al. 2004; Aguerri et al. 2004). 

The aim of the present paper is to classify as many galaxies as it is feasible in the Coma
cluster, increasing the number of members detected from velocity data by 
considering object morphology, apparent size, and surface brightness. In subsequent
papers of this series, we will study 
the resulting member galaxies populations. These can be considered from two aspects: 
in-cluster variations and cluster-to-cluster comparisons. For this second approach we 
chose the Virgo cluster because of the superior quality and completeness of the  
BST2 data. 

In the present paper, Sect. 2 summarizes the technical aspects of the classification,  
describes its output catalogs and presents an independent test of the validity of 
the morphological membership in the Coma cluster, i.e. the determination of
the membership based on galaxy morphology, apparent size, and surface brightness. Section 3 presents our
essential data: the catalog, the histograms of the various Hubble-types for
Coma and the Virgo comparison sample, and a table showing the sample size of each
morphological type in both clusters.
We also introduce an auxiliary catalog of V magnitudes and U$-$V colors 
derived from the data by Terlevich et al. (2001) (TCB01) and others. 
Section 4 briefly summarizes the results and gives 
some highlights of a later paper(s) in this series.

\begin{figure}
 \resizebox{\hsize}{!}{\includegraphics{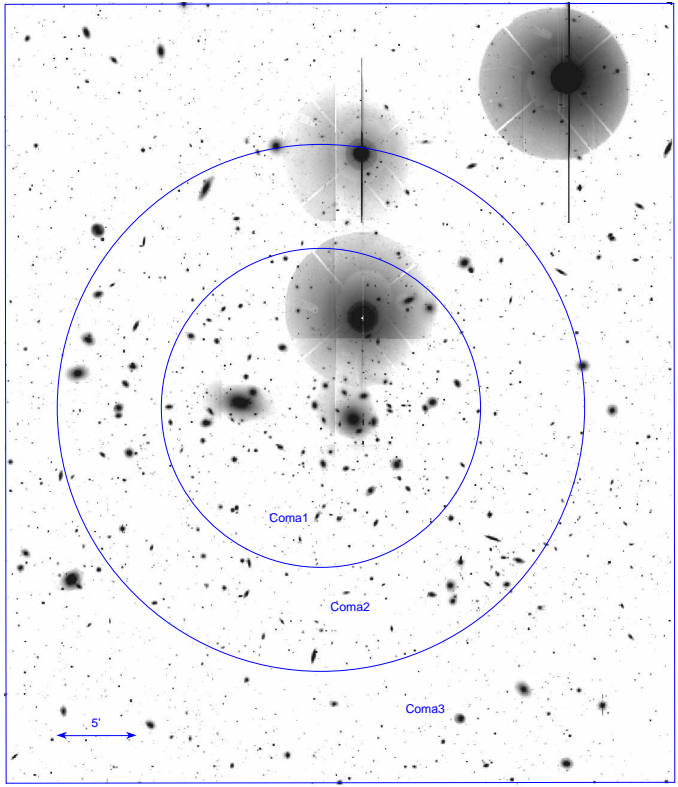}}
\caption[ ]{
B-band image of the studied portion of the Coma clusters. The image is highly
binned ($16 \times 16$) for display purposes. The 3 zones used in the discussion 
of radial population changes are marked. North is at the top, east to the left.
}
\label{studiedfield}
\end{figure}

\section{Morphological classification and membership}
\subsection{The used samples} 
The initial sample contains 1154 objects of the GMP catalog (Godwin et al. 1983)
falling in the field of the CFH12K frames: the cut-off limit of the GMP is 
at $\mathrm{B_{GMP}}=21.0$, this magnitude being defined within the
$\mu_\mathrm{b}=26.5$ isophote. 
We added one galaxy, omitted in the GMP list, at X=879, Y=579 in the GMP coordinate 
system. We had to discard 17 stars and a number of objects that could not be 
classified for various reasons: CCD defects, vicinity of a bright star, 
severe blend, or truncation by a CCD edge. 

We assume that the distance modulus of Coma is $35.0$ mag (Colless, 2000). 
On the other hand, we found the $\mathrm{B_{GMP}}$ magnitudes to be 0.25 mag fainter in 
the mean than the usual B$_t^0$ total corrected magnitudes of the RC2 or RC3 systems. 
This is derived from a comparison with the HYPERLEDA data base involving 38 bright 
E+S0 objects, with a scatter of 0.20 mag. The shift between the two systems may vary
with magnitude and also Hubble-types. We adopt, however, the simple expression 
$\mathrm{M_B}=\mathrm{B_{GMP}}-35.25$ to transform the GMP catalog B magnitudes into 
absolute corrected magnitudes. 

The Virgo comparison sample was derived from the catalog of 
BST2 and is intended to cover the same domain of absolute magnitude as the 
Coma sample. We adopt 
a distance modulus of 31.05 mag for the cluster (Ferrarese et al. 1996).    
A similar comparison as above led to $\mathrm{B_t^0=B_{BST}}-0.10$ for 152 objects 
of all types ($\sigma=0.25$). Then absolute magnitudes were obtained from the expression 
$\mathrm{M_B=B_{BST}}-31.15$. We selected in the BST catalog the objects noted 
M, for members, with $\mathrm{B_{BST}} < 16.90$.  We then rejected objects with 
the unusual types, BCD, or Amorphu. We had to suppress duplicates noted by the authors with the 
symbols $or$ or $/$, the first named cell being retained. 
The final comparison Virgo sample numbers 514 galaxies.    

\subsection{Morphological classification}
The deep frames from the CFH12K camera offer a convenient way to study Coma cluster 
galaxies in a relatively large field, i.e. 42x55$\arcmin$, and at $1\arcsec$ FWHM 
seeing (Adami et al. 2006). Exposure times are 7200 s in $B$ and $\sim 4600$ in $V$.
One of us (RM) inspected the objects in the GMP catalog to determine the
Hubble-types.  We
tried to follow the classification system adopted in a classical work on the 
Virgo cluster (Sandage \& Binggeli, 1984; Sandage et al., 1985a; id., 1985b; 
Binggeli et al., 1985 or BST2), but the refinements introduced there for bright 
spirals could not be retained. 
In the early-type range, 
the types E, SA0, SB0, and E/S0 with the eventual addition of p for peculiar, and 
of c for compact (Ec exclusively) were considered. 
For spirals, the types SAa, SAb, SAc and their counterparts 
(SBa, SBb, SBc) for barred galaxies were used. 
Intermediate types like Sab, Sbc... have been disregarded. 
We often selected the type SA0/a to describe an SA0 (or SB0)
seemingly surrounded by tightly wound spiral arms or, as usual, doubtful objects 
at the S0-Sa limit. 
Dwarf early-type galaxies are classified dE, dS0, Ec, where the latter class 
is for compact objects of the M32 type. 
For late-type dwarfs Sm and Im were used, plus a single Sdm. 
Part of our Sm objects might have possibly been classified Sd by others.

In classifying faint objects, it was found useful to look at both the B and V  
frames simultaneously because the characteristics features of spirals, i.e. massive dust and 
young stars systems, have more contrast in B. On the other hand, far-away E/S0 look 
distinctly brighter in V, while cluster dE look alike in both bands.    
To view the images a color look-up table of the MIDAS system was used, which 
displays isophotal contours well. This helps to detect diskyness in E or S0, 
bars in S0 or Sa, and, departures of symmetry revealing of 
blurred spiral structure for distant objects. 
 
Several controls were made of our classification:
\begin{itemize} 
\item Since two widely overlapping frames are needed to cover the studied portion
of the cluster, independent classifications can be obtained for a subsample of 64 
bright objects. Our first attempt showed poor self-consistency, but a more
stringent application of the available criteria brought notable improvements.   

\item We classified 27 Virgo cluster spiral galaxies from DSS2 frames, having
practically the same scale  in kpc/pix as the CFH12K frames, although with 
a slightly broader PSF. Our types disagree with the Binggeli et al. catalog by one 
stage of the spiral sequence in 3 cases, and by two stages in one case. There are two
contradictions as regards the occurrence of a bar. 

\item We inspected F475W (i.e. approximately B) HST images taken with 
the Advanced Camera for Surveys for a dozen of dwarf objects, confirming  
the previously adopted types in all cases. 

\item The morphological classification of Coma cluster galaxies has previously been 
restricted to bright objects. Andreon \& Davoust (1997) compare all classifications 
then available. 
A subsample of 56 relatively bright objects ($\mathrm{B}<16.5$ mag) is available for
comparison of the present work with Andreon et al. (1997). There are only a few 
contradictions: 2 changes of E against S0, 3 of Sa against S0. 
We also missed two bars seen in the earlier work. 
It is gratifying to find such good agreement in the separation of S0 from E, since 
in Andreon (loc. cit.), this was based on a quantitative photometric analysis.

\item Among the 247 galaxies classified by Dressler (1980) in a Coma cluster 
field of 2.1 sq.deg., we found coincidences with 124 objects on our list of cluster 
members in the much smaller field of the CFHT camera. 
Dressler's objects are relatively bright galaxies: B$>17$ objects are very scarce. 
There are notable differences between the present system of classification and the
one of Dressler. In our work, the adopted type tends to be a later one because 
we shifted 29\% of Dressler's E to S0, 33\% of his S0 to S0/a, half his S0/a to Sa, 
most of his very rare Sa to Sb. We only agree for the late spirals. 
These systematic differences are not too surprising, since the morphological 
classification is a matter of personal application of rather loose criteria.   
Morphologies in the seminal work by Dressler (1980) shows systematics with
respect not only to ours, but also to Andreon et al. (1996, 1997) and others 
(see Andreon \& Davoust 1997).  
\end{itemize}  

\subsection{Membership from morphological appearance?}
We attempted to classify fainter objects, up to the limit of the GMP 
catalog. An increasingly large proportion of these are distant giant galaxies, 
while the cluster members are dwarfs.  One purpose of the classification then is  
to identify members that are too faint to have been included in the available redshift data.

\subsubsection{Morphological criteria for faint-object membership}
It is possible to guess whether a galaxy is a member of Coma or lies in the background 
thanks to a number of rather uncertain criteria. 
First, dwarf spirals do not exist, according to Sandage et al. (1985a).  
Typical spirals are then expected to disappear from the Coma population at a 
$\mathrm{B}\approx 19$ mag. Bright, far-away spirals appearing in the same magnitude 
range as faint spiral members are scarce because they lie at the bright end of the LF. 
Fainter and farther background spirals lose many of the features that allow a 
detailed classification in the 3 stages Sa, Sb, Sc. They can still be recognized, 
however, because the open spiral arms lead to an often asymmetric 
twist of the outer isophotal contours. Also, the abundance of dust results in a
characteristic asymmetry along the minor axis. Therefore the morphology  
together with apparent size and brightness, gives a sufficient indication of the distance 
of a spiral to classify it as a member or background object. 
 
Dwarfs galaxies in Coma should appear around B=18 and become the unique
cluster population for $\mathrm{B}>19$. 
There is a morphological similarity between dE or dS0 objects in Coma and distant 
giant E/S0's, but the latter appear more sharply bounded and generally have a smaller linear 
diameter. In ambiguous cases, we did not hesitate to bring colors to the rescue as 
a final test, since distant E/S0 galaxies are generally strongly K-reddened.   

\subsubsection{Morphology assigned membership against redshift data}
The $cz$ values remain the primary indicators of cluster membership, even if we tried 
to supplement these by considering forms, apparent size 
and eventually colors. Our two main sources of $cz$ data were a query in NED and 
a compilation provided by Biviano (see Adami et al. 2005). We also consulted original 
sources (Colless \& Dunn, 1996; Castander et al., 2001; Mobasher et al., 2001; 
Rines et al., 2003). 
The objects were classified without knowing their redshift, which were incorporated
afterwards in to the tables of results.    
Generally, there was agreement between the morphology-based and redshift-based
membership assignations with only a few conflicts. The redshift-based indication was 
generally preferred. We should, however, mention the cases of 
GMP2943, 3038, 3131, 3258, 3336, 3353, and 3615 with strongly conflicting $cz$ in the 
available sources: the $cz$ value compatible with the morphology was then adopted
in these 7 cases. 
For the 9 objects GMP2353, 2605, 4364, 2412, 2422, 3370, 3563, 4108, and 4424,
the available $cz$ seemed incompatible with the morphology, apparent size, and color,
the redshift-based assignation was then discarded. 
Of special interest are galaxies with $cz < 4000 \mathrm{km/s}$, the usual inward 
limit for membership, none of these shows the expected aspect of a foreground galaxy!
They were located in the cluster (GMP2293, 2633, 3262, 3275, 3633, 3780) or the 
background (GMP2538, 3583, 3642). According to the redshift analysis by 
Biviano et al. (1996), the close pair GMP3262-75 is part of a group in 
front of the cluster.  

We emphasize that
morphological-based memberships disagree with spectroscopical memberships 9 times,
as much as various authors strongly disagree on the redshift of our galaxies
(7 occurrences, listed above). The good agreement between morphology-based
and spectroscopical-based membership testifies that our morphological
membership can be trusted, at least for galaxies with reshift. The next section
shows the good agreement for all galaxies, including
those without a known redshift, by means of a statistical argument: the luminosity
function.

\begin{figure}
 \resizebox{\hsize}{!}{\includegraphics{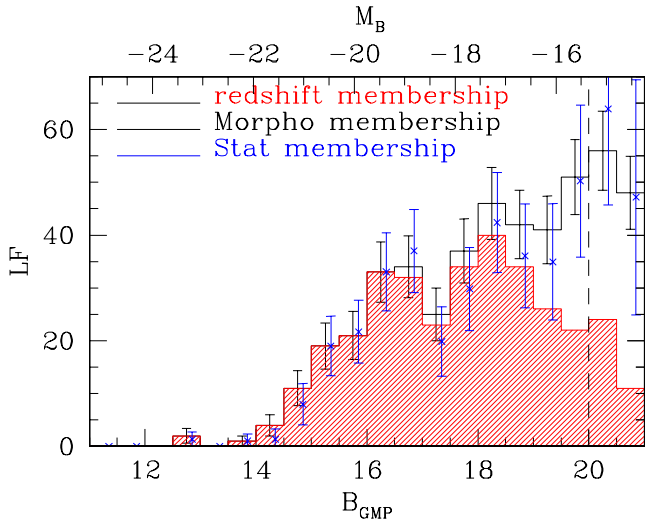}}
\caption[ ]{Coma cluster luminosity function (LF) as derived by three
methods. The (red) hashed histogram only considers spectroscopically
known members with increasing incompleteness below B=18 mag. The solid (black) 
histogram shows the LF of the galaxies estimated to be member from their morphological
appearance. The blue points mark the LF derived by statistical
subtraction of background galaxies, slightly offset for clarity. The standard 
deviation of the observed number of galaxies (accounting for background, when relevant)
is plotted with error bars. The top abscissae marks the absolute magnitude and 
the vertical dashed line indicates the GMP completeness limit. Note the excellent 
agreement between the morphology- and statistically derived memberships.  
}
\label{LFgmp}
\end{figure}

\begin{figure}
\centerline{\includegraphics[width=70mm]{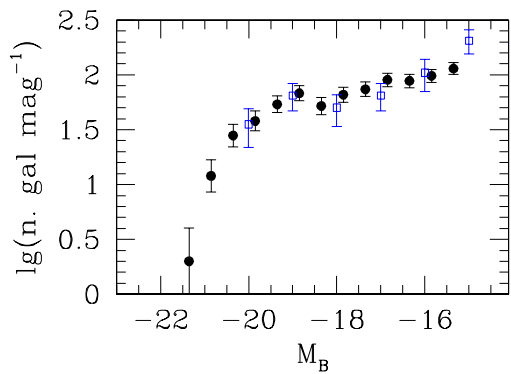}}
\caption[ ]{Coma-cluster luminosity function (LF) of the galaxies estimated to be 
members from their morphological appearance (solid dots) and as
derived by statistical
subtraction of background galaxies from fully independent and CCD photometry and
using a control field uncontaminated by cluster galaxies (blue open points, from
Andreon \& Cuillandre 2002). The morphology- and statistically derived 
LFs again well agree, showing that we can trust the morphological
membership even below the spectroscopic limit. }
\label{lf_morpho_stat}
\end{figure}

\subsubsection{A second test of morphology-based memberships}

Figure \ref{LFgmp} shows the B band luminosity functions of Coma galaxies in the
studied region built in three different ways. First, we considered spectroscopically 
confirmed members.
The spectroscopic sample is incomplete at $\mathrm{B}>18$, due to the lack of available redshifts 
for fainter and fainter objects. 
Second, we considered the sample formed by galaxies that are estimated to be members 
from their morphological appearance. 
Brigher than the spectroscopic completness limit,
morphology-based and spectroscopic-based LF well agree (Fig 2).  

Finally, we performed the usual statistical
subtraction of background/foreground galaxies using a control field. 
Still using the GMP catalog we adopt 
an annulus between 1.0 and 1.3 degrees as control field. 
We only considered galaxies bluer than B$-$R=2.0 mag, in order to reduce the
contribution of background objects without affecting the members contribution,
because galaxies at the Coma distance have B$-$R $< 1.8$ mag. This annulus is too
close to the cluster, i.e. is contaminated by its galaxies. The main effect of
taking too near a control field is to reduce the luminosity function by a
multiplicative factor without affecting its shape too much (Paolillo et al.
2001). This multiplicative factor can be found by asking that the statistically
derived number of Coma galaxies brighter than B=16.75 mag be equal to the number 
of spectroscopy-based memberships, and it turns out to be 1.3. 
The LF of Coma galaxies computed from the morphological membership 
(solid line histogram) appears to be indistinguishable from the one obtained by the 
statistical method (blue data points). The above holds true both 
at the bright end (B $< 16.75$ mag), where by construction the integral of the two LFs 
are forced to be equal, but also at B$ > 16.75$ mag where the two derivations are 
independent. Towards the faint end of the 
sample, where spectroscopic information is typically available for 50 \%  or less of 
the galaxies, the morphological approach possibly recovers all the members 
counted in the statistical approach. 
The above comparison, summarized in Fig. 2, shows that our extended list of 
members is statistically correct. It has the same level of completeness as the GMP 
catalog, which is the source of both our morphology and of the two samples involved in 
the statistical derivation of the LF. 
According to the authors, the GMP is nearly complete up to $\mathrm{B_{GMP}}=20$ mag, 
but is only 50\% complete one magnitude fainter.  

A similar experiment is repeated in Fig. 3, but using independent 
deeper CCD photometry (Andreon \& Cuillandre 2002), no color-selection at all, and a control field 
unaffected by Coma galaxies, and leads to the same
conclusion. Estimating the membership from the morphological appearance is feasible, 
not only for nearby clusters such as Virgo, but even at the distance of Coma, although the 
criteria cannot be the same. In this comparision, the compared
sample is complete down to some two magnitude deeper. The found good agreement 
between morphology-derived and statistically-derived LFs 
extends our conclusion that morphological membership 
can be trusted to galaxies without a known redshift, 
down to $M_B\sim 15.0$ mag.

\begin{table}
\caption[ ]{Sample sizes of morphological populations for the Virgo sample and 4 
Coma samples. 
}
\begin{flushleft}  
\begin{tabular}{llllll}
\hline
Data &  Virgo & Coma0 & Coma1 & Coma2 & Coma3 \\       
\hline
cD &    0 &    2 &     2 &     0 &     0    \\ 
E &	 20 &	37 &	15 &	10 &	12   \\ 
E/S0 &  6 &	11 &	6 &	3 &	2    \\ 
S0 &	 40 &	61 &	29 &	16 &	16   \\ 
S0/a &  7 &	43 &	15 &	20 &	8    \\ 
Sa &	 26 &	56 &	21 &	15 &	20   \\ 
Sb &	 13 &	24 &	4 &	11 &	9   \\ 
Sc &	 62 &	7  &	0 &	1 &	6    \\ 
\hline
giants & 174 &  241 &  93 &    75 &    73    \\  
\hline
Ec &	 1 &	13 &   10 &    2 &     1     \\    
dE &	 224 &  129 &  49 &    45 &    35    \\
dS0 &   28 &	43 &   12 &    15 &    16    \\
dlat &  87 &	47 &   21 &    12 &    14    \\
\hline 
dwarfs & 340 &  232 &  92 &    74 &    66    \\ 
\hline
\hline
\end{tabular}
\end{flushleft}
The Table gives the counts of the various Hubble types in the 5 samples. 
dlat refers the late-type dwarfs Sd, Sdm, Sm, Im. 
Partial sums are also given for the ''giant" and 
''dwarf" galaxies respectively. 
\end{table}

\begin{figure}
\centerline{\includegraphics[width=70mm]{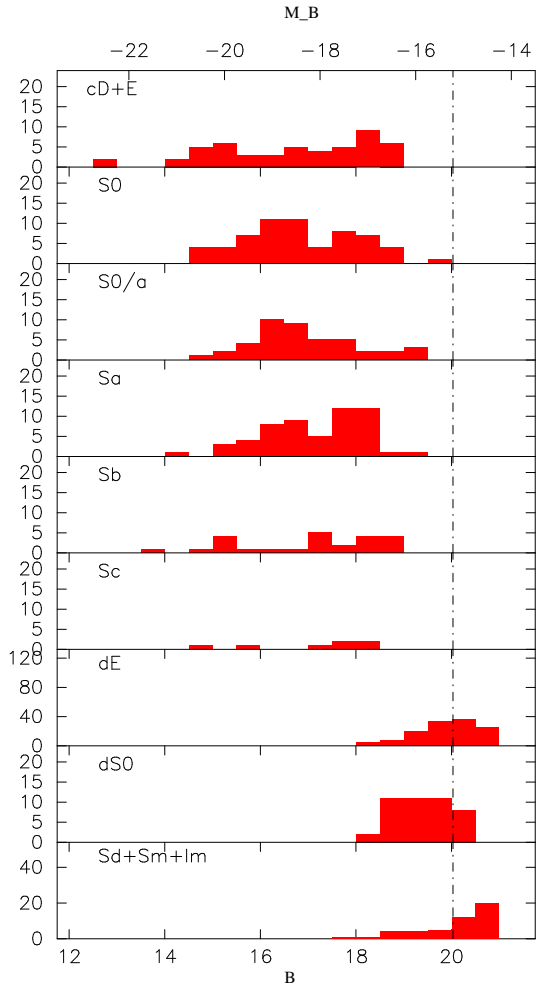}}
\caption[ ]{Coma-cluster luminosity histograms of each morphological
type. Upper abscissae scale: absolute B magnitude $\mathrm{M_B}$. 
Lower abscissae scale: GMP B magnitude $\mathrm{B_{GMP}}$ 
The vertical dashed line marks the limit of completeness of the GMP catalog at 
$\mathrm{B_{GMP}}=20$.
Types cD, E and E/S0 are added together and also late-type dwarfs 
Sd+Sm+Im. Note the important change in the Y-scale for the dwarfs dE et 
Sd+Sm+Im (also dlate). 
}
\label{comaHtyp_hist}
\end{figure}

\begin{figure}
\centerline{\includegraphics[width=70mm]{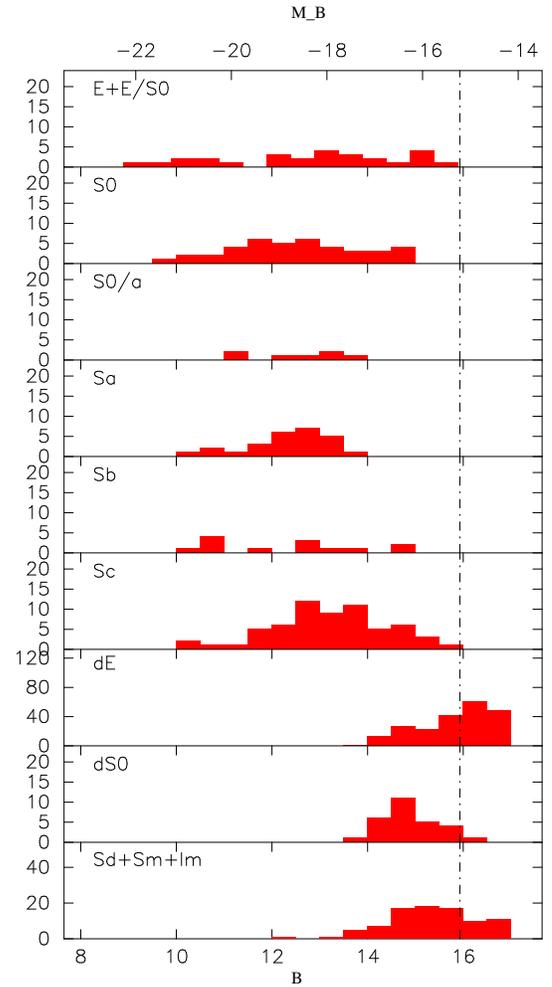}}
\caption[]{Virgo-cluster luminosity histograms  of each morphological type. 
The conventions are the same as for the Coma histograms, but for the lower 
magnitude scale taken from the BST2 catalog, truncated at the same luminosity 
as our Coma sample. 
}
\label{virgoHtyp_hist}
\end{figure}

\section{Presentation of the data}
\begin{enumerate}
\item {\em The morphology and membership catalogs}: 
Table E1, available at CDS and at the link
http://www.brera.inaf.it/utenti/andreon/E1, lists 1155 entries containing:
(1) the number $igmp$ of the object in the GMP catalog ($igmp=9999$ for the 
first entry, a galaxy missing in the GMP); (2) and (3) GMP coordinates; 
(4) GMP B-magnitude; (5) GMP B$-$R color; (6) adopted redshift among available
measurements, see Sect 2.3.2; (7) estimated Hubble type; (8) adopted membership, 
$in$ for members or $far$ for background objects;
(9) Notes: $ble$ for blend i.e. strongly overlapping objects; 
$dlo$ doubtful location; $pip$ possibly interacting pair; $sin$ strongly inclined; 
$tfc$ too faint to classify; $tpb$ technical problem (CCD defects, bright star 
too close, image cut by CCD edge, etc.).
Incomplete information are replaced by dots. For instance ... deals with an 
unclassified object; S.. is a spiral of indeterminate class; S.0 might be either 
an SA0 or SB0. 

In Table E1, 473 galaxies are classified as Coma members, of which 343 have 
measured redshifts. Membership is, by its nature, probabilistic: i.e. there is 
(almost) never absolute certainty that a galaxy is a cluster member, yet 
galaxies have been put into one of the two classes: member or non-member. 
Therefore, any analysis
of our morphological membership flags should take into account that 
a handful of galaxies may have a wrong membership and 
consequently may have inappropriate properties for Coma members, for 
example to be an outlier in a color-magnitude diagram. This nuisance is 
positively offset by the advantage of an individual correct classification for 
the very large majority of the sample, because galaxy properties (say color 
distributions) can be studied without the need to account for background 
contamination and to do so with good precision, because the abundant background 
population is almost completely filtered out by our morphological membership. An 
example is given in our analysis (Andreon \& Michard, 2008, in preparation) of 
the color-magnitude relation, using an outlier-tolerant Student $t$ distribution 
in place of a Gaussian.

\item {\em Histograms of luminosity for the various Hubble-type.}  
Figures 4 and 5 show the LF of each morphological type in Coma
and Virgo, respectively. The order of
presentation is the order of the Hubble sequence from cD, normal E, through 
S0, and spiral stages Sa to Sc, and then the dwarfs dE's, dS0's, dlate 
(i.e. Sd, Sm, Im). The compact dwarfs Ec are not entered here because of the 
uncertainty of their classification. 
For the first time, the morphological classification of Coma 
galaxies reaches magnitudes that are faint enough 
to observe complete samples of the giant types, E, S0, and spiral stages. 
The corresponding histograms are limited at both the bright and faint ends of 
the 
distributions. We also observed a significant part of the dwarf types 
populations, whose brightest objects are at $\mathrm{M_B}=-17$ and extend to 
$\mathrm{M_B}=-14.25$, our sample being complete about 1 mag brighter. 

\item {\em Sample sizes}.
Table 1 collects the number of the Coma (and Virgo for comparison) members of each
each morphological type: our full sample termed Coma0 and three 
subsamples of increasing
distances to the cluster center, termed Coma1, Coma2, and Coma3. 
Coma1 refers to the core of the cluster, i.e. a circle of $10\arcmin$ 
radius around the conventional GMP center; Coma2 is a concentric ring of 16.67 
outer radius; Coma3 is the rest of the field around Coma2, as shown in Fig. 1. 
Our Coma0 sample is larger than the Virgo sample for giant types earlier than Sc.

\item {\em The spiral stages in Coma and Virgo}.
The population balance between the 3 spiral stages drastically changes between the 
two clusters, as seen in Table 1 or Figs. 4 and 5. 
In our Coma sample there are twice more Sa than in  
Virgo, but there are 12 times less Sc. 
The combined population Sb+Sc is thrice less important in Coma than in Virgo. 
Andreon \& Michard (2008, in preparation) will further quantify the subject.
Our results are in line with the the morphology-density 
relation but with an hitherto 
undetected difference in the segregation of the various spiral stages, because 
in previous works (e.g. Dressler 1980; Andreon, 1996; Whitmore et al., 1993), 
the distinction between 
the successive spiral stages could not be made and samples were limited to 
bright galaxies.

\item {\em Auxillary U$-$V catalogs}
The above morphological data has been completed by V$_t$, U$-$V catalogs 
from Terlevich et al. (2001,
TCB01), who give U and V magnitudes 
within several apertures, 8.8 to $26\arcsec$.  The 
color listed in our Table E2 is the mean color, averaged over all 
the available apertures within a 
maximal radius increasing with galaxy luminosity to account for the larger
size of brighter galaxies. Due to the integrated nature of aperture colors
and the large size of the adopted apertures,
color gradients present in regions of low surface brightness outside the
largest aperture minimally affect the measured value, because we have
already integrated most of the galaxy flux. 
For E+S0 giants, the U$-$V here derived agree  
with the Bower et al. (1992) (BLE92) values,  fully corrected for K-effect 
and galactic absorption, and also with the data for 20 objects in common with 
Prugniel \& Simien (1996). For 43 objects we find a mean 
$\mathrm{(U-V)_{Bow}-(U-V)_{us}}=-.008$ mag with $\sigma=0.041$ mag. 
Random errors assumed equal for both sets have $\sigma=0.03$ mag in the 
range V$_t^0<16$ mag. 
No such comparisons are feasible for fainter objects.

The ``total" magnitudes listed in Table E2 were computed in two
steps. First, we started from the aperture magnitude in the largest
aperture used for the color measurement. For dwarf galaxies, which are 
smaller than the $13\arcsec$ aperture, this is also the final color.
Brighter objects, in practice giant galaxies, are truncated even in 
the largest aperture available from Terlevich et al. (2001), 26 arcsec. 
We obtained corrections for this effect from comparisons 
with V$_{26}$ of Lobo et al. (1997) and  the V$_t^0$ in BLE92. 
For the cD, we adopted the R magnitudes of Andreon et al. (1996) with a 
shift 
V$-$R =0.60 mag. 
Our final V$_t$ scale coincides for giant E+S0 galaxies (after excluding
these two cD's) with the one of BLE92 (mean difference of -0.013 mag
and $\sigma=0.23$ mag).  This may suggest a mean error of 0.15 mag in both sets. 
Our sample of Coma members with 
Hubble-types and U$-$V colors numbers 314 galaxies, 60 \% of which are giants 
and 
40 \% dwarfs. 
This is a subsample extracted from the list of GMP members defined above: the
faint objects measured in TCB01 but not in the GMP were examined but rejected 
as probably in the background. 
Table E2, available at CDS and at the link 
http://www.brera.inaf.it/utenti/andreon/E2, lists for the 314 members:
(1) Terlevich et al. identification number; (2) the number $igmp$ of the object in the GMP 
catalog; (4) and (5) GMP coordinates; (6) aperture corrected 
$V_t$; (7) $V_t$ error, rounded
to two digits from $V$ mag values listed in Terlevich et al.; (8) aperture
corrected $U-V$; (9) $U-V$ error, rounded
to two digits from $U-V$ mag values listed in Terlevich et al.; (10) 
adopted redshift among available
measurements; (11) estimated Hubble type; (12) membership.

\end{enumerate}

\section{Summary of results and highlights of forthcoming papers}

A catalog of morphological types were obtained for 1155 objects in the
Coma clusterfield, and memberships estimated from their appearance, i.e.
morphology, surface brightness, rarely color, and, aftewards, redshifts. 
Four hundred seventy-three
cluster members were identified, and it has been shown 
that our memberships can be trusted both when individual
redshifts are available (Sect. 2.3.2), and when membership is only known 
in a statistical sense (Sect. 2.3.3).
We obtained from the literature 
V$_t$/U$-$V data for 2/3 of our Coma members and corrected them for
aperture effects.

For the first time, the morphological classification of Coma galaxies
reaches magnitudes faint enough to observe the full magnitude range of 
the
giant types, E, S0, and spiral stages. Tables with morphological types,
membership, and U-V colors are given.
Our sample is the
richest environment where membership and morphology for complete samples 
down to faint magnitudes ($M_B\sim-15$ mag) are available,
thereby enlarging the baseline of environmental studies.

Some new results  will be presented in detail in future papers.
a) Dwarf galaxies are more abundant in Virgo than in Coma.
b) It appears that late-type dwarfs in Coma are less numerous, less bright, 
and less active in star formation than in Virgo; c) We derived
LF parameters of the giant galaxies in Coma, 
distinguishing between various Hubble-types and covering the full luminosity range 
for each type. This has not been possible up to now, except for the Virgo 
cluster in the classical work of Sandage et al. (1985), which we  
follow closely in our study. The LF of dwarfs can also be studied paying the 
necessary 
attention to sample incompleteness.

\begin{acknowledgements}
We thank Drs. Adami, Biviano, and Durret for their help in gathering useful 
data. 
\end{acknowledgements}

{}


\begin{thebibliography}{}
\bibitem[2005]{adami}
Adami C., Biviano A, Durret F. et al., 2005, A\&A 443, 17
\bibitem[2006]{adami}
Adami C., Picat J.P., Savine C. et al., 2006, A\&A 451, 1159
\bibitem[2004]{aguerri}
Aguerri J.A.L., Iglesias-Pramo, J., Vilchez, J.M. et al. 2004, AJ 127, 1344
\bibitem[Andreon(1996)]{1996A&A...314..763A} 
Andreon, S.\ 1996, A\&A, 314, 763 
\bibitem[Andreon \& Davoust(1997)]{1997A&A...319..747A} 
Andreon, S., \& Davoust, E.\ 1997, A\&A, 319, 747 
\bibitem[1996]{andreon}
Andreon S., Davoust E., Michard R. et al., 1996, A\&AS 116, 429
\bibitem[1997]{andreon}
Andreon S., Davoust E., Poulain P., 1997, A\&AS 126, 67
\bibitem[2002]{andreon}
Andreon S., Cuillandre J.C., 2002, ApJ 569, 144
\bibitem[Binggeli et al.(2000)]{2000eaa..bookE1822B} 
Binggeli, B. \& Huchra, J.\ 2000, Encyclopedia of Astronomy and Astrophysics 
\bibitem[1985]{binggeli}
Binggeli B., Sandage A., Tammann G.A., 1985, AJ 90, 1681 (BST2)
\bibitem[Binggeli et al.(1993)]{1993A&AS...98..275B} 
Binggeli, B., Popescu, C.~C., \& Tammann, G.~A.\ 1993, A\&AS, 98, 275 
\bibitem[1996]{biviano}
Biviano A., Durret F., Gerbal D. et al., 1996, A\&A 311, 95
\bibitem[1992]{bower}
Bower R.G., Lusey J.R., Ellis R.S. 1992, MNRAS 254, 601 (BLE92)
\bibitem[2000]{buta}
Buta R., 2000,  Encyclopedia of Astronomy and Astrophysics 
\bibitem[2001]{castander}
Castander F.J., Nichol F.C., Merrelli A. et al., 2001, AJ 121, 2331
\bibitem[1996]{colless}
Colless M., Dunn A.M., 1996, ApJ 438, 435
\bibitem[2000]{colless}
Colless M., 2000, Encycl. Astr. Astrophys. 2000 
\bibitem[1980a]{dressler}
Dressler A., 1980a, ApJS 42, 565
\bibitem[1980b]{dressler}
Dressler A., 1980b, ApJ 236, 351
\bibitem[Ferrarese et al.(1996)]{1996ApJ...464..568F} 
Ferrarese, L., et al.\ 1996, ApJ, 464, 568 
\bibitem[1983]{godwin}
Godwin J.G., Metcalfe N., Peach J.V., 1983, MNRAS 202, 113 (GMP)
\bibitem[2004]{guttierez}
Guttierez C.M., Trujillo I., Aguerri, J.A.L. 2004, ApJ 602, 664
\bibitem[Jorgensen \& Franx(1994)]{1994ApJ...433..553J} 
Jorgensen, I., \& Franx, M.\ 1994, ApJ, 433, 553 
\bibitem[1997]{lobo}
Lobo C., Biviano A., Durret F. et al.,1997, A\&AS 122, 409
\bibitem[1995]{michard}
Michard R., 1995, ApL\&C 31, 187
\bibitem[2001]{mobasher}
Mobasher B., Bridges T.J., Carter D. et al., 2001, ApJS 137, 279
\bibitem[2001]{paolillo}
Paolillo, M., Andreon, S., Longo, G. et al., 2001, A\&A 367, 549 
\bibitem[1996]{prugniel}
Prugniel P., Simien F., 1996, A\&A 309, 749
\bibitem[2003]{rines}
Rines K., Geller M.J., Kurz M.J. et al., 2003, AJ 126, 2152
\bibitem[1984]{sandage}
Sandage A., Binggeli B., 1984, AJ 89, 919 (Paper III)
\bibitem[1985a]{sandage}
Sandage A., Binggeli B, Tammann G.A., 1985a, AJ 90, 395 (Paper IV)
\bibitem[1985b]{sandage}
Sandage A., Binggeli B, Tammann G.A., 1985b, AJ 90, 1759 (Paper V)
\bibitem[2005]{sandage}
Sandage A.,2005, ARA\&A 43, 581
\bibitem[2001]{terlevich}
Terlevich A.I., Caldwell R., Bower R.G., 2001, MNRAS 326, 1547 (TCB01)
\bibitem[1993]{Whitmore}
Whitmore B.C., Gilmore D.M.,Jones C., 1993, ApJ 407, 489
\end{thebibliography}
\end{document}